\documentclass[12pt]{article}

\usepackage{graphicx}

\def\rf#1{(\ref{eq:#1})}
\def\lab#1{\label{eq:#1}}

\def\br{\begin{eqnarray}}
\def\er{\end{eqnarray}}
\def\be{\begin{equation}}
\def\ee{\end{equation}}
\def\({\left(}
\def\){\right)}

\def\rlx{\relax\leavevmode}
\def\IR{\rlx\hbox{\rm I\kern-.18em R}}
\def\vp{\varphi}
\def\u2{\mid u\mid^2}

\def\IZ{\rlx\hbox{\sf Z\kern-.4em Z}}
\def\IR{\rlx\hbox{\rm I\kern-.18em R}}
\def\IC{\rlx\hbox{\,$\inbar\kern-.3em{\rm C}$}}

\begin{document}

\begin{titlepage}
\vspace*{-1cm}

\vskip 3cm

\vspace{.2in}
\begin{center}
{\large\bf Self-dual Hopfions}
\end{center}

\vspace{.3cm}

\begin{center}
L. A. Ferreira$^{(1)}$ and A.C. Ris\'erio do Bonfim$^{(2)}$

\vspace{.3 in}
\small

\par \vskip .2in \noindent
$^{(1)}$Instituto de F\'\i sica de S\~ao Carlos; IFSC/USP;\\
Universidade de S\~ao Paulo  \\ 
Avenida Trabalhador S\~ao-carlense 400, Caixa Postal 369\\ 
CEP 13560-970, S\~ao Carlos-SP, Brazil\\

\vspace{.3 in}
\small

\par \vskip .2in \noindent
$^{(2)}$Instituto de F\'\i sica Te\'orica; IFT/UNESP;\\
Universidade Estadual Paulista\\
Rua Dr. Bento Teobaldo Ferraz 271 - Bl. II - Barra Funda\\
CEP 01140-070, S\~ao Paulo-SP, Brazil

\normalsize
\end{center}

\vspace{.5in}

\begin{abstract}

We construct static and time-dependent exact soliton solutions with
non-trivial Hopf topological charge for a field theory in $3+1$
dimensions with the target space being the two dimensional sphere
$S^2$. The model considered is a reduction of the so-called extended
Skyrme-Faddeev theory by the removal of the quadratic term in
derivatives of the fields. The solutions are constructed using an
ansatz based on the conformal and target space symmetries. The
solutions are said self-dual because they solve  first order
differential equations which together with some conditions on the coupling
constants, imply the second order equations of motion. The solutions
belong to a sub-sector of the theory with an
infinite number of local conserved currents.  The
equation for the profile function of the 
ansatz corresponds to the Bogomolny equation for the sine-Gordon model.

\end{abstract} 
\end{titlepage}

\section{Introduction}
\label{sec:intro}
\setcounter{equation}{0}

Two important aspects in the study of non-linear phenomena are the
identification of the symmetries and the relevant degrees of freedom
in a given regime of the dynamics. In many cases solitons connect
those two aspects in a very special way. On one hand, the appearance
of solitons is related to a high degree of symmetries and so of
conservation laws. On the other hand, solitons are in many cases the
proper relevant degrees of freedom. In low dimensional phenomena the
soliton theory is quite well developed, with the prototype being the
sine-Gordon model. That is an integrable field theory which in the weak
coupling regime is described by the elementary excitations of a scalar
field, and at strong coupling the natural degrees of freedom become 
 the solitons with their dynamics being governed by the massive Thirring
model \cite{coleman}. The fact that a unique quantum field theory can
be described 
by two different classical Lagrangians is a remarkable fact, and it is
believed that similar phenomena occur in higher dimensions, with the
electromagnetic duality in gauge theories being the most important
example \cite{duality}. 

Soliton theory in dimensions higher than two is far from being well
understood, even at the classical level. One way of addressing the
problem is to look for the equivalent of the structures responsible
for the conservation laws in two dimensional soliton theories,
i.e. the zero curvature condition or Lax-Zakharov-Shabat equation
\cite{lax}. The 
results obtained so far show that loop spaces play a crucial role in
that approach. It has been shown that the generalization of the
Lax-Zakharov-Shabat equation to higher dimensions can be best
formulated as the zero curvature condition for a connection on
generalized loop spaces \cite{afs,afs-review}. Conservation laws are then
obtained in way very similar to that of two dimensions. Most of the
known soliton theories fit into that 
scheme. However, it is not clear how the full loop space structure
works to render them solvable, i.e. to provide a method for
constructing exact solutions.   

There are two main contexts in which solitons appear in field
theories. As the solutions of the classical equations of motion, like
the instantons and magnetic monopoles in non-abelian gauge theories, 
or then as solutions of low energy effective actions like the
skyrmions and hopfions \cite{skyrme,faddeev-niemi,mantonbook}. The purpose of
this paper is to study solitons   
belonging to that second class. We consider a field theory 
described by the Lagrangian 
\be
{\cal L} = -\frac{1}{e^2} \, H_{\mu\nu}^2 + \frac{\beta}{2}\,
\left(\partial_{\mu} {\vec n}\cdot\partial^{\mu} {\vec n}\right)^2
\lab{action}
\ee
where ${\vec n}$ is a triplet of scalar fields taking values on the
two dimensional sphere $S^2$, i.e. ${\vec n}^2=1$, 
$H_{\mu\nu}$ is the pull back of the area form on $S^2$,  $e^2$ and
$\beta$ are dimensionless coupling constants. We shall consider the
theory \rf{action} in four dimensional 
Minkowski  space-time and also in three and four dimensional
Euclidean spaces.  Notice that the second term in \rf{action} is
quartic in time derivatives and so it breaks the positivity of the
energy. However, as we explain below, we shall be considering sectors
of the theory \rf{action} where the energy
is indeed positive definite. 

Using the
stereographic projection one can parametrize $S^2$ with a complex
scalar field $u$, related to ${\vec n}$ as
\be
{\vec n} = \frac{1}{1+\u2}\,\(u+u^*,-i\(u-u^*\),\u2 -1\)
\lab{ndef}
\ee
In addition, we shall parametrize $u$ as
\be
u = \sqrt{\frac{1-g}{g}}\; e^{i\,\theta}
\lab{gthetadef}
\ee
where $g$ and $\theta$ are real scalar fields, with $0\leq g\leq 1$,
and $0\leq
\theta \leq 2\pi$,  playing the role of Darboux variables for the
two-form $H_{\mu\nu}$. We then have
\be
H_{\mu\nu}\equiv 
{\vec n}\cdot\(\partial_{\mu}{\vec n} \wedge 
\partial_{\nu}{\vec n}\)
=-2i\frac{\(\partial_{\mu} u\partial_{\nu} u^* - 
 \partial_{\nu} u \partial_{\mu} u^*\)}{\(1+\u2\)^2}  
=  2\,\left[\partial_{\mu}g\,\partial_{\nu}\theta-
\partial_{\nu}g\,\partial_{\mu}\theta\right]  
\lab{hdef} 
\ee
and
\be
\left(\partial_{\mu} {\vec
  n}\cdot \partial^{\mu} {\vec
  n}\right)=4\,\frac{\partial_{\mu}u\;\partial^{\mu}u^*}{\(1+\u2\)^2}
=\frac{\(\partial_{\mu}g\)^2}{g\(1-g\)}
+4\,g\(1-g\)\,\(\partial_{\mu}\theta\)^2  
\lab{dn2}
\ee
The Lagrangian \rf{action} is then written as 
\be
{\cal L}= 
\frac{8}{e^2}\left[ 
\frac{\(\partial_{\mu}u\)^2\(\partial_{\nu}u^*\)^2}{\(1+\u2\)^4}+
\(\beta\,e^2-1\)\,\frac{\(\partial_{\mu}u\;\partial^{\mu}u^*\)^2}{\(1+\u2\)^4}
\right]
\lab{actionu}
\ee

The theory \rf{action} is invariant under the global $SO(3)$ rotations
on the target space $S^2$, and also invariant under the conformal
groups $SO(4,2)$ or $SO(5,1)$, depending if we consider Minkowski or
Euclidean four dimensional space-time respectively. Those two
symmetries play a crucial 
role in this paper since the soliton solutions will be constructed,
following \cite{babelon}, 
with an ansatz invariant under the combined action of the $U(1)$ subgroup
of $SO(3)$ responsible for the phase transfomation $u\rightarrow
e^{i\alpha}\,u$, and three special commuting $U(1)$ subgroups of the
conformal group. Similar ans\"atze were used in \cite{afz,laf1,laf2}.  

The Lagrangian \rf{action} is part of the Wilsonian low energy effective
Lagrangian for the pure (without matter) $SU(2)$ Yang-Mills theory
\cite{gies}, if 
one assumes that 
the Cho-Faddeev-Niemi decomposition of the gauge potentials
\cite{chofn} holds true 
at low energies. The missing term is the quadratic operator, 
$M^2\,\partial_{\mu} {\vec n}\cdot \partial^{\mu} {\vec n}$, with $M$
being a mass parameter generated by the so-called dimensional
transmutation. Even though that term breaks the conformal symmetry, 
exact vortex solutions have been constructed for such 
model \cite{vortexlaf}. The solutions belong to a  special
sector defined by particular relations among the 
coupling constants and a constraint that leads to an infinite number
of conservation laws. In addition, static Hopf topological solitons
have been constructed numerically for the theory with the $M^2$ term
and outside that special sector 
\cite{fst}. For the case of the Faddeev-Niemi
model \cite{faddeev-niemi} (without the second term in
\rf{action} but with the quadratic term mentioned above), numerical
solutions have  been calculated in 
\cite{sutcliffe,hietarinta}. The
solitons we calculate in this paper have a free parameter which allow
to rescale their size and rate of time evolution. That is a
consequence of the conformal invariance of the equations of motion and
it may play a role in the study of the solutions of the theory with the
$M^2$ term.

The special property of theory \rf{action} is that it admits a Bogomolny type
equation, i.e. a first order differential equation implying the second
order equations of motion. That is given by \cite{afs}
\be
\(\partial_{\mu}\,u\)^2=0
\lab{bogomolny}
\ee
together with the following relation among the coupling constants
\be
\beta\, e^2=1
\lab{criticalcoupling}
\ee
An interesting point is that for the ansatz configurations we consider, 
the equation \rf{bogomolny} becomes the Bogomolny equation for the
sine-Gordon model. Therefore, the self-dual soliton solutions of
\rf{action} at the critical point \rf{criticalcoupling}, are built
from the static one-soliton of the sine-Gordon model. 
The exact vortex solutions constructed in \cite{vortexlaf} belong to
the sector where $\beta\,e^2=1$ and they satisfy \rf{bogomolny}.  

The equation \rf{bogomolny} also plays an important role in the
integrability properties of the model. The theory \rf{action}  has three
conserved currents associated to the $SO(3)$ target space
symmetry. However, for the sub-model satisfying the
constraint \rf{bogomolny},  in
addition to the Euler-Lagrange equations associated to \rf{action}, 
the number of conserved currents become 
infinite. Indeed, the equations of motion for \rf{action}  are 
\be
\(1+\u2\)\, \partial^{\mu}{\cal K}_{\mu}-2\,u^{*}\,{\cal K}_{\mu}\,
\partial^{\mu} u=0
\lab{eom}
\ee
and its complex conjugate, where
\be
{\cal K}_{\mu}\equiv \frac{ 
\(\beta\,e^2-1\)\,\(\partial_{\nu}u\,\partial^{\nu}u^{*}\)\,\partial_{\mu}
u+\(\partial_{\nu}u\)^2 \partial_{\mu}u^{*}}{\(1+\u2\)^2}
\lab{kdef}
\ee
If one imposes \rf{bogomolny} as a constraint it follows that the
currents \cite{afs,babelon}
\be
J_{\mu}\equiv \frac{\delta G}{\delta u}\, {\cal K}^c_{\mu} - 
\frac{\delta G}{\delta u^{*}}\, {{\cal K}^c_{\mu}}^{*}
\lab{conscurr} 
\ee
are conserved for any functional $G$ of $u$ and $u^{*}$, but not of
their derivatives, and where ${\cal K}^c_{\mu}$ is obtained
from \rf{kdef} by imposing \rf{bogomolny}, i.e.
\be
{\cal K}^c_{\mu}\equiv \frac{ 
\(\beta\,e^2-1\)\,\(\partial_{\nu}u\,\partial^{\nu}u^{*}\)\,\partial_{\mu}
u}{\(1+\u2\)^2}
\lab{kcdef}
\ee
The conservation of \rf{conscurr} follows from the fact that ${\cal
  K}_{\mu}\partial^{\mu}u^{*}$ is real, ${\cal K}^c_{\mu}
\partial^{\mu}u=0$, and from the equations of motion which now read
$\partial^{\mu}{\cal K}^c_{\mu}=0$. Notice that all the currents 
\rf{conscurr}  vanish at the critical point \rf{criticalcoupling}. 

The theory \rf{action} possesses another sub-model with an infinite
number of conserved currents defined by the constraint
\be
\partial_{\mu}u\,\partial^{\mu}u^*=0
\lab{bogomolnycomplex}
\ee
Consider the currents
\be
{\tilde J}_{\mu}\equiv \frac{\delta {\tilde G}}{\delta u}\,
{\tilde{\cal K}}^c_{\mu} -  
\frac{\delta {\tilde G}}{\delta u^{*}}\, \({\tilde{\cal K}}^c_{\mu}\)^{*}
\lab{conscurr2} 
\ee
where ${\tilde G}$ is a functional of $u$ and $u^*$ but not of their
derivatives, and ${\tilde{\cal K}}^c_{\mu}$ is obtained from \rf{kdef}
by imposing \rf{bogomolnycomplex}, i.e.
\be
{\tilde{\cal K}}^c_{\mu}\equiv \frac{ 
\(\partial_{\nu}u\)^2 \partial_{\mu}u^{*}}{\(1+\u2\)^2}
\lab{ktcdef}
\ee
Similarly, the conservation of \rf{conscurr2} follows from the fact
that ${\tilde{\cal K}}^c_{\mu}\partial^{\mu}u^{*}$ is real,
${\tilde{\cal K}}^c_{\mu} \partial^{\mu}u=0$, and from the equations
of motion which now read $\partial^{\mu}{\tilde{\cal K}}^c_{\mu}=0$.

The importance of the constraints \rf{bogomolny} and
\rf{bogomolnycomplex} is twofold. Besides leading to an infinite number of
conserved currents as we showed above, they also make the energy
positive definite, as we now explain. Notice that, by using \rf{hdef},
\rf{dn2} and 
\rf{actionu}, the Lagrangian \rf{action}  can be written in three
different ways as 
\be
{\cal L}=  \frac{\(\beta\,e^2-1\)}{e^2}\,{\cal L}_1
+ \frac{1}{e^2}\, {\cal L}_2 = 
 -\frac{1}{e^2} \, H_{\mu\nu}^2 + \beta\, {\cal L}_1 = 
 \frac{\(\beta\,e^2-1\)}{e^2} \, H_{\mu\nu}^2 + \beta\, {\cal L}_2
\lab{3ways}
\ee
where $H_{\mu\nu}$ is given in \rf{hdef}, and where we have defined
\be
{\cal L}_1 = 8\,
\frac{\(\partial_{\mu}u\,\partial^{\mu}u^*\)^2}{\(1+\u2\)^4}\qquad\qquad\qquad
{\cal L}_2 = 8\,
\frac{\(\partial_{\mu}u\)^2\,\(\partial_{\nu}u^*\)^2}{\(1+\u2\)^4}
\ee
The Legendre transform of ${\cal L}_1$ is 
\be
{\cal H}_1\equiv {\dot u}\,\frac{\delta {\cal L}_1}{\delta {\dot u}} + 
{\dot u}^*\,\frac{\delta {\cal L}_1}{\delta {\dot u}^*} - {\cal L}_1 = 
24\,\frac{\left[\mid{\dot u}\mid^2+
\frac{1}{3}\,{\vec \nabla}u\cdot {\vec\nabla}u^*\right]
\left[\mid{\dot u}\mid^2-
{\vec \nabla}u\cdot {\vec\nabla}u^*\right]}{\(1+\u2\)^4}
\lab{h1}
\ee
where ${\dot u}$ denotes the $x^0$-derivative of $u$, and ${\vec
  \nabla}u$ its spatial gradient. 
Notice that ${\cal H}_1=0$  when \rf{bogomolnycomplex} holds true. On the
other hand the Legendre transform of ${\cal L}_2$ is 
\be
{\cal H}_2\equiv 
24\,\frac{\({\vec \nabla}u\)^2\,\( {\vec \nabla}u^*\)^2}{\(1+\u2\)^4}
\,\left[F^2 - \(\frac{2}{3}\)^2\right]
\lab{h2}
\ee
where we have denoted
\be 
\frac{{\dot u}^2}{\({\vec \nabla}u\)^2}\equiv \frac{1}{3}+
F\, e^{i\,\Phi}
\lab{fphidef}
\ee
with $F>0$ and $0\leq\Phi\leq 2\pi$, being functions of the space-time
coordinates.  
Therefore, ${\cal H}_2$ vanishes on a circle on the $\left[\frac{{\dot
    u}^2}{\({\vec \nabla}u\)^2}\right]$-complex plane, with radius equal to
$\frac{2}{3}$ and center at  $\frac{{\dot
    u}^2}{\({\vec \nabla}u\)^2}=\frac{1}{3}$. 
Notice that the constraint \rf{bogomolny}, for the Minkowski case,
lies on that circle and 
corresponds to $\Phi =0$ and $F = \frac{2}{3}$. In fact,
\rf{bogomolny}  is the only Lorentz invariant condition that lies on
that circle.   

The Legendre transform of the term $\left[- H_{\mu\nu}^2\right]$ is
always positive definite. Therefore, from \rf{3ways}, \rf{h1} and
\rf{h2} one obtains the following result. 

{\bf Theorem 1.} {\em The Hamiltonian density 
${\cal H}$ associated to \rf{action} is positive definite when: 
\begin{itemize}
\item The constraint \rf{bogomolny} holds true and
  $\frac{\(\beta\,e^2-1\)}{e^2}<0$.
\item The constraint \rf{bogomolnycomplex} holds true and $e^2>0$.
\end{itemize} }

A more careful analysis, exploring \rf{h1}, \rf{h2} and the three ways
of writing the Lagrangian as in \rf{3ways}, leads to the following result. 

{\bf Theorem 2.} {\em The Hamiltonian density 
${\cal H}$ associated to \rf{action} is positive definite when: 
\begin{itemize}
\item $e^2<0$, $\beta <0$ and $F\leq \frac{2}{3}$. 
\item $e^2>0$, $\beta <0$ and $F\leq \frac{2}{3}$ or $\mid{\dot
    u}\mid^2\leq {\vec \nabla}u\cdot {\vec\nabla}u^*$, 
\end{itemize} 
with $F$ defined in \rf{fphidef}.}

There are sectors where the Hamiltonian is positive definite for
$\beta>0$, but they are not physically interesting since they exclude
the static configurations.  

The papers is organized as follows. In section \ref{sec:ansatz} we
constructed the ansatz based on the conformal symmetry and the target
space phase transformation $u\rightarrow e^{i\,\alpha}\,u$. Using that
ansatz  we construct the exact hopfion self-dual solutions. In
sub-section \ref{sec:static} we construct the static version of those
hopfions. In sections \ref{sec:hopf} and \ref{sec:energy} we calculate
the Hopf topological charges and energies of those solutions
respectively. Our results are also valid on Euclidean four dimensional
space-time and also for a version of the theory \rf{action} which
possesses positive definite energy. Those results are shown in section
\ref{sec:other}.  A summary of the results as well as the conclusions
are presented in section \ref{sec:conclusions}. The  technical
results are given in the appendices.

\section{The ansatz}
\label{sec:ansatz}
\setcounter{equation}{0}

We now construct exact soliton solutions for the theory \rf{action} in
the integrable sub-sectors defined by the constraints \rf{bogomolny}
and \rf{bogomolnycomplex}, using a special ansatz based on the
conformal and internal symmetries.   
In order to implement the ansatz we have to work with special
coordinates. Three of them are chosen to parametrize curves generated
by three commuting  $U(1)$ subgroups of the conformal group $SO(4,2)$
 of Minkowski  space-time. The
fourth one, denoted $z$ in what follows,  is taken to parametrize
curves orthogonal to those three 
(see \cite{babelon} for details). In
Minkowski space-time they are related to the Cartesian coordinates
$x^{\mu}$, $\mu =0,1,2,3$, as\footnote{Notice they correspond to the
  coordinates used in \cite{laf1} with the change $y\rightarrow
  \frac{1-z}{z}$, and to the toroidal coordinates used in \cite{afz}
  for $\zeta = 0$ and $z={\rm tanh}^2 \eta$.}  
\br
x^0&=& \frac{r_0}{p}\, \sin \zeta \qquad\qquad\qquad\;\,
x^1= \frac{r_0}{p}\, \sqrt{z}\; \cos\varphi \nonumber\\
x^3&=& \frac{r_0}{p}\, \sqrt{1-z}\; \sin\xi \qquad\quad
x^2= \frac{r_0}{p}\, \sqrt{z}\; \sin\varphi 
\lab{minkcoord}
\er
where $r_0$ is a constant with dimension of length, and 
\be
p=\cos\zeta-\sqrt{1-z}\;\cos\xi 
\ee
The range of those coordinates are: $0\leq z\leq 1$, $0\leq \varphi\,
,\, \xi\leq 2\,\pi$, and $0\leq \zeta \leq \pi$. Notice that
$\(\zeta,z,\xi,\varphi\)$ and $\(\zeta+\pi,z,\xi+\pi,\varphi+\pi\)$,
describe the same point in space, and that is why the range of
$\zeta$ is restricted. 
The Minkowski metric (with signature $\(+,-,-,-\)$), in those
coordinates, becomes  
\br
ds^2 = \(\frac{r_0}{p}\)^2\;\left[ d\zeta^2 -
  \frac{dz^2}{4\,z\,\(1-z\)}-\(1-z\)\,d\xi^2 - z\, d\varphi^2\right]
\nonumber
\er
The ansatz corresponds to those field configurations where the scalar
fields introduced in \rf{gthetadef} are of the form
\be
g\equiv g\(z\) \qquad \qquad \theta\equiv m_1\, \xi +m_2\, \varphi+m_3\,
\zeta
\lab{mink-ansatz}
\ee
where $m_i$, $i=1,2,3$, are integers satisfying the condition
$m_1+m_2+m_3= \mbox{\rm even integer}$. That is required in order for
the complex scalar field $u$ to be single valued, since
$\(\zeta=0,z,\xi,\varphi\)$ and
$\(\zeta=\pi,z,\xi+\pi,\varphi+\pi\)$, correspond to the same point
in space. 

Replacing the ansatz \rf{mink-ansatz}  (see \rf{gthetadef}) into the
equations of motion \rf{eom}  
associated to \rf{action}, one gets 
that the equation for $\theta$ is automatically satisfied and 
that the profile function $g\(z\)$ has to satisfy
\br
\partial_z\left[\left[\(\frac{z\(1-z\)}{g\(1-g\)}\,g^{\prime}\)^2 + 
\gamma\,\Lambda_M\right]g^{\prime}\right]
+\frac{\(1-2g\) g\(1-g\)}{2\,\(z\(1-z\)\)^2} 
\left[\(\frac{z\(1-z\)}{g\(1-g\)}\,g^{\prime}\)^4-\Lambda_M^2\right]=0
\lab{geq}
\er
where $g^{\prime}\equiv \partial_z g$, and 
\be
\Lambda_M \equiv z\,m_1^2+\(1-z\)\,m_2^2-z\(1-z\)\,m_3^2
\lab{lambdadef}
\ee
and
\be
\gamma = \(1-\frac{2}{\beta\,e^2}\)
\lab{gammadef}
\ee
Notice that the equation \rf{geq} is invariant under the following two
$\IZ_2$ transformations
\be
g \leftrightarrow 1-g
\ee
and
\be
z\leftrightarrow 1-z \qquad \qquad\qquad m_1^2 \leftrightarrow m_2^2
\ee

Now replacing the ansatz \rf{mink-ansatz} into the constraints
\rf{bogomolny} and \rf{bogomolnycomplex} one gets that
\be
\(\partial_{\mu}\,u\)^2=0 \qquad \qquad\leftrightarrow \qquad\qquad 
\left[\frac{z\(1-z\)}{g\(1-g\)}\, g^{\prime}\right]^2=\Lambda_M
\lab{bogomolny2} 
\ee
and
\be
\partial_{\mu}\,u\, \partial^{\mu}\,u^*=0 \qquad \qquad\leftrightarrow
\qquad\qquad  
\left[\frac{z\(1-z\)}{g\(1-g\)}\, g^{\prime}\right]^2=\,-\,\Lambda_M
\lab{bogomolnycomplex2} 
\ee
The equation  \rf{bogomolny2}  obviously  has acceptable
solutions only if 
$\Lambda_M \geq 0$  on the interval $0\leq z\leq 1$. Analogously,
\rf{bogomolnycomplex2}  has  solutions only for $\Lambda_M \leq 0$. The
results of appendix \ref{sec:appendixlambdam} show that, on the
interval $0\leq z\leq 1$,  
\be
\Lambda_M \geq 0 \qquad\qquad {\rm if}\qquad\qquad m_1,\, m_2\neq 0 
\qquad {\rm and}\qquad \(\mid m_1\mid + \mid m_2\mid\)^2>m_3^2 
\lab{lambdacond1}
\ee
and
\be
\Lambda_M\leq 0 \qquad\quad {\rm if} \qquad\quad m_1=m_2=0\qquad
\mbox{\rm and so } \qquad \Lambda_M=-z\(1-z\)\,m_3^2
\lab{lambdacond2}
\ee
If $g$ satisfies \rf{bogomolny2} or \rf{bogomolnycomplex2} then
\rf{geq} reduces to  
\be
\(\gamma \pm 1\)\,\partial_z\left[\Lambda_M\,g^{\prime}\right]=0
\lab{geqreduced}
\ee
where the factor $\(\gamma + 1\)$ corresponds to \rf{bogomolny2} and
$\(\gamma - 1\)$ to \rf{bogomolnycomplex2}. For $\gamma \neq \pm 1$
the equation \rf{geqreduced} implies that $g^{\prime}\sim
1/\Lambda_M$.  If $m_3\neq 0$, $\Lambda_M$ is quadratic in $z$ and so
of the form $\Lambda_M \sim \(z-z_+\)\(z-z_-\)$. Therefore, $g\sim
c_1\,\ln\left[\frac{z-z_+}{z-z_-}\right]+c_2$, and so it does not
satisfy \rf{bogomolny2} or \rf{bogomolnycomplex2}, even for
$c_1=0$. If $m_3=0$ then $\Lambda_M$ is linear in $z$, and so $g$ is
still a logarithm, and consequently not a solution of  \rf
{bogomolny2} or \rf{bogomolnycomplex2}. The only remaining possibility
is to have $\Lambda_M= {\rm constant}$, and so $m_3=0$ and $m_1^2=m_2^2$. For
the case of \rf{bogomolnycomplex2} we then have $\Lambda_M=0$ due to
\rf{lambdacond2}, and so the only acceptable solution is $g= {
\rm constant}$, which  implies ${\vec n}= {\rm constant}$. For the
case of \rf{bogomolny2} we then have $\Lambda_M=m_2^2$, and so
\rf{geqreduced} implies that $g$ is linear in $z$. But \rf{bogomolny2}
then requires that $g\(1-g\)\sim z\(1-z\)$. The only way for $g$ to be
linear in $z$ and satisfy $g\(1-g\)=\alpha\, z\(1-z\)$, is for
$\alpha=1$, and then either $g=z$ or $g=1-z$. In such case we need
$\Lambda_M=1$. Therefore, the only solutions of \rf{bogomolny2} and
\rf{geqreduced}, for $\gamma\neq -1$, and so $\beta\, e^2\neq 1$, are
\be
m_1^2=m_2^2=1;\qquad m_3=0 \qquad\qquad{\rm and}\qquad\qquad g=z\qquad {\rm
  or}\qquad g=1-z 
\lab{nicesolforg}
\ee
In terms of the ${\vec n}$ field (see \rf{ndef} and \rf{gthetadef})
such solutions read
\be
{\vec n} =
\(2\,\sqrt{z\(1-z\)}\,\cos\(\varepsilon_1\,\xi+\varepsilon_2\,\varphi\), 
2\,\sqrt{z\(1-z\)}\,\sin\(\varepsilon_1\,\xi+\varepsilon_2\,\varphi\),
\varepsilon_3\,\(1-2\,z\)\)
\lab{nicesolforn}
\ee
with $\varepsilon_i=\pm 1$, $i=1,2,3$, and $m_1=\varepsilon_1$,
$m_2=\varepsilon_2$, $\varepsilon_3=1$ for $g=z$ and
$\varepsilon_3=-1$ for $g=1-z$.  Notice that although such
solutions are $\zeta$-independent they are time dependent. 

From the above discussion we also have that the only solutions of
\rf{bogomolnycomplex2} and 
\rf{geqreduced}, for $\gamma\neq 1$, and so finite $\beta\, e^2$, are
those were the field ${\vec n}$ is constant. 

Therefore, except for the solution \rf{nicesolforg}-\rf{nicesolforn},
the only physically interesting solutions of \rf{action}, within the
ansatz \rf{gthetadef}-\rf{mink-ansatz}, are those at the critical
points $\gamma=\pm1$, 
which solve the constraints \rf{bogomolny2} and
\rf{bogomolnycomplex2}.  In that sense the conditions  
\rf{bogomolny} and   \rf{bogomolnycomplex} 
 constitute  Bogomolny type equations for the
model \rf{action} at the critical points $\beta\,e^2=1$ and
$\beta\,e^2\rightarrow \infty$ respectively. 

The integration of the equations \rf{bogomolny2} and
\rf{bogomolnycomplex2} is quite simple. We can write them as
\be
\frac{d\;}{d z}\left[\ln\frac{g}{1-g}\right]= \varepsilon\, \frac{\sqrt{\mid
    \Lambda_M\mid}}{z\(1-z\)}\; , \qquad \qquad\qquad \varepsilon=\pm 1
\lab{bogomolny3}
\ee
and so
\be
g=\frac{ e^{\varepsilon\, w}}{1+e^{\varepsilon\, w}}
\lab{geqw}
\ee
with
\be
w= \int_{\kappa}^z
dz^{\prime}\;\frac{\sqrt{\mid\Lambda_M\mid}}{z^{\prime}\(1-z^{\prime}\)}\; , 
\qquad \qquad {\rm with} \qquad 0< \kappa < 1
\lab{wdef}
\ee
where the integration constant was encoded into 
$\kappa$. Notice that the integrand in \rf{wdef} is positive, and so $w$
is negative for $0<z<\kappa$ and it is positive for $\kappa < z < 1$,
and vanishes at $z=\kappa$. In addition, $w$ is a monotonically
increasing function of $z$.  

We point out that the equations  \rf{bogomolny2} and \rf{bogomolnycomplex2} 
correspond in fact to the static Bogomolny equation for the sine-Gordon
model. Indeed, if one defines 
\be
 g = \sin^2\(\frac{\phi}{4}\)\; ,  \qquad \qquad 0\leq \phi \leq 2\,\pi
\ee
then the equation \rf{bogomolny3} becomes 
\be
\frac{d\,\phi}{d\,w}=\varepsilon\, \sin \frac{\phi}{2}
\ee
which implies the second order static sine-Gordon equation,
$\frac{d^2\,\phi}{d\,w^2}=\frac{1}{4} \sin \phi$. Therefore, the
solutions  \rf{geqw} correspond to the soliton ($\varepsilon =1$) and
anti-soliton ($\varepsilon =-1$) solutions of the sine-Gordon model. 

For the solutions of \rf{bogomolnycomplex2} we have $\Lambda_M$ given
by \rf{lambdacond2} and so \rf{wdef} gives
\be
w_{(-)}=2\mid m_3\mid\left[{\rm ArcTan}\;\sqrt{\frac{z}{1-z}}-
{\rm ArcTan}\;\sqrt{\frac{\kappa}{1-\kappa}}\;\;\right]
\ee
where the index ${(-)}$ refer to the $w$ solutions for the case
\rf{bogomolnycomplex2} and \rf{lambdacond2}. 
The field $g$ is given by \rf{geqw} and so using \rf{ndef} and
\rf{gthetadef} one gets the corresponding solution for the ${\vec n}$
field as
\be
{\vec n}=\(\frac{\cos\(m_3\,\zeta\)}{\cosh\frac{w_{(-)}}{2}},
\frac{\sin\(m_3\,\zeta\)}{\cosh\frac{w_{(-)}}{2}},-\varepsilon\,
\tanh\frac{w_{(-)}}{2}\)
\lab{bogosoln-1}
\ee
Notice that as $z$ varies from $0$ to $1$, $w_{(-)}$ varies on finite
range and it never diverges. Therefore, the $n_3$ component never
reaches the values $1$ or $-1$.  

That differs from the behavior of the solutions for the case
\rf{bogomolny2} and \rf{lambdacond1}. In that case $m_1$ and $m_2$ do not
vanish, and from \rf{lambdadef} we have
\br
\frac{\sqrt{\mid\Lambda_M\mid}}{z\(1-z\)} \sim \left\{\begin{array}{ll}
\frac{\mid m_2\mid}{z} & \quad {\rm for}\qquad z\sim 0\\
\frac{\mid m_1\mid}{1-z}& \quad {\rm for}\qquad z\sim 1
\end{array}\right.
\er
Therefore, $w$, given in \rf{wdef}, is logarithmically divergent for
$z\sim 0$ and $z\sim 
1$, and so $w \rightarrow -\infty$ for $z\rightarrow 0$, and $w
\rightarrow +\infty$ for $z\rightarrow 1$. Consequently, one observes
from \rf{geqw} that the profile function $g$ satisfy the boundary
conditions 
\br
g\(0\)=0 &\qquad& \qquad g\(1\)=1 \qquad \qquad {\rm for} \qquad
\varepsilon =1\nonumber\\
g\(0\)=1 &\qquad& \qquad g\(1\)=0 \qquad \qquad {\rm for} \qquad
\varepsilon =-1
\er
and $g\(z=\kappa\)= \frac{1}{2} $ in both cases. Clearly, $g$ is a
monotomic function of $w$ and so of $z$. Therefore, we have
\be
{\vec n}\(z=0\)=\(0,0,\varepsilon\)\qquad\qquad\qquad 
{\vec n}\(z=1\)=\(0,0,-\varepsilon\)
\ee

In fact, the solution for the ${\vec n}$ field in that case is
\be
{\vec
  n}=\(\frac{\cos\(m_1\,\xi+m_2\,\varphi+m_3\,\zeta\)}{\cosh\frac{w}{2}},
\frac{\sin\(m_1\,\xi+m_2\,\varphi+m_3\,\zeta\)}{\cosh\frac{w}{2}},
-\varepsilon\,\tanh\frac{w}{2}\)
\lab{bogogensoln}
\ee
with
\be
\beta\,e^2=1
\qquad\qquad m_1,\, m_2\neq 0 
\qquad {\rm and}\qquad \(\mid m_1\mid + \mid m_2\mid\)^2>m_3^2 
\lab{restrictions}
\ee
Notice that the solutions \rf{bogosoln-1} and \rf{bogogensoln} depend
on a free parameter $\kappa$, with $0< \kappa <1$. That parameter
appeared in \rf{wdef} as an integration constant. It determines where
$w$ vanishes, i.e. $w\(z=\kappa\)=0$. Therefore, from  \rf{bogosoln-1}
and \rf{bogogensoln} on observes that $z=\kappa$ is the point where
the third component of ${\vec n}$ vanishes. Such parameter is
associated to a symmetry of the equations  \rf{bogomolny2} and
\rf{bogomolnycomplex2}. Indeed, they were written as \rf{bogomolny3},
which in its turn can be cast as
$\frac{d\,}{d\,w}\left[\ln\frac{g}{1-g}\right]=\varepsilon$. Therefore,
it is invariant under the translations $w\rightarrow w+{\rm const.}$,
which amounts to the freedom encoded into $\kappa$. 

Notice that, from \rf{wdef} one has 
\be
w= \mid m\mid\,\ln\left[\frac{z}{1-z}\,\;\frac{1-\kappa}{\kappa}\right]
\qquad\qquad {\rm for} \qquad
\mid m_1\mid=\mid m_2\mid\equiv \mid m\mid \quad m_3=0 
\ee
Therefore the solution \rf{bogogensoln} coincides with
\rf{nicesolforn} for $m_1^2=m_2^2=1$, $m_3=0$ and $\kappa =1/2$.

\subsection{Static solutions}
\label{sec:static}

The equations of motion \rf{eom}, as well as the action associated to
\rf{action}, are conformally invariant in a four dimensional
space-time, i.e.,  in $3+1$ or $4+0$ dimensions.  That is what allowed
us to use the ansatz \rf{gthetadef}-\rf{mink-ansatz}, since it is
invariant under the direct product of commuting $U(1)$ subgroups of
the conformal group with the internal $U(1)$ subgroup generated by the phase
transformations $u\rightarrow e^{i\,\alpha}\,u$. The constraints
\rf{bogomolny} and \rf{bogomolnycomplex} are conformally invariant in
a space-time of any number of dimensions. Consequently, we can use an 
ansatz based on the conformal group to solve those constraints in any
dimension. We shall do that to construct static solutions of the
theory \rf{action} in three spatial dimensions. Notice that the
conditions 
\be
\beta\,e^2=1\qquad \qquad\qquad \partial_0u=0\qquad\qquad\qquad 
{\vec \nabla}u\cdot {\vec \nabla}u=0
\lab{staticbogomolny}
\ee
or
\be
\beta\,e^2\rightarrow \infty\qquad\qquad
\qquad \partial_0u=0\qquad\qquad\qquad {\vec \nabla}u\cdot {\vec
  \nabla}u^*=0
\lab{staticbogomolnycomplex}
\ee
where ${\vec \nabla}u$ is the spatial gradient of $u$, 
are sufficient conditions for the equations of motion \rf{eom} to be
satisfied. 

The existence of static solutions of the theory \rf{action} seems to
be against Derrick's theorem
\cite{derrick}, since the energy scales as $E\rightarrow
\frac{1}{\lambda}\,E$ as $x^i\rightarrow \lambda \, x^i$. However, 
for the integrable sub-sectors where the constraints
\rf{staticbogomolny} or \rf{staticbogomolnycomplex} holds true the
Derrick's theorem does not 
apply. The reason for that may be related to the fact that very
probably there is not a Lagrangian such that the equation \rf{eom} and
the constraint \rf{bogomolny}, or
\rf{bogomolnycomplex}, can  be derived as the
corresponding Euler-Lagrange equations. 

 Therefore, 
we shall also calculate three dimensional static solutions using the toroidal
coordinates given by
\be
x^1= \frac{r_0}{{\tilde p}}\, \sqrt{z}\; \cos\varphi \qquad \quad 
x^2= \frac{r_0}{{\tilde p}}\, \sqrt{z}\; \sin\varphi \qquad \quad
x^3= \frac{r_0}{{\tilde p}}\, \sqrt{1-z}\; \sin\xi 
\lab{3dcoord}
\ee
with
\be
{\tilde p} = 1-\sqrt{1-z}\;\cos\xi
\ee
Notice they are obtained from the coordinates \rf{minkcoord}  by
setting $\zeta=0$. The metric is given by
\br
ds^2 = \(\frac{r_0}{{\tilde p}}\)^2\;\left[
  \frac{dz^2}{4\,z\,\(1-z\)}+\(1-z\)\,d\xi^2 + z\, d\varphi^2\right]
\nonumber
\er
The ansatz in this case corresponds to (see \rf{gthetadef}) 
\be
g\equiv g\(z\) \qquad \qquad \theta\equiv m_1\, \xi +m_2\, \varphi
\lab{3d-ansatz}
\ee
with $m_1$ and $m_2$ being integers. Replacing that into the
constraints \rf{staticbogomolny} or \rf{staticbogomolnycomplex} one gets
\be
{\vec \nabla}u\cdot {\vec \nabla}u=0 \qquad\qquad \leftrightarrow 
\qquad \qquad
\left[\frac{z\(1-z\)}{g\(1-g\)}\, g^{\prime}\right]^2=\Lambda_S
\lab{bogomolny2static} 
\ee
and 
\be
{\vec \nabla}u\cdot {\vec \nabla}u^*=0 \qquad\qquad \leftrightarrow 
\qquad \qquad
\left[\frac{z\(1-z\)}{g\(1-g\)}\, g^{\prime}\right]^2=-\Lambda_S
\lab{bogomolnycomplex2static} 
\ee
where 
\be
\Lambda_S \equiv z\,m_1^2+\(1-z\)\,m_2^2
\ee
Since, $\Lambda_S\geq 0$ on the interval $0\leq z\leq 1$, for any $m_1$
and $m_2$, it is clear that there are no non-trivial solutions for
\rf{staticbogomolnycomplex} inside such ansatz. 

The integration of \rf{bogomolny2static} can be done following
\rf{bogomolny3}-\rf{wdef} to give
\be
g=\frac{ e^{\varepsilon\, w}}{1+e^{\varepsilon\, w}}
\lab{geqwstatic}
\ee
with
\be
w= \int_{\kappa}^z
dz^{\prime}\;\frac{\sqrt{\Lambda_S}}{z^{\prime}\(1-z^{\prime}\)}\; , 
\qquad \qquad {\rm with} \qquad 0< \kappa < 1
\lab{wdefstatic}
\ee
The corresponding solutions for the ${\vec n}$ field are 
\be
{\vec
  n}=\(\frac{\cos\(m_1\,\xi+m_2\,\varphi\)}{\cosh\frac{w}{2}},
\frac{\sin\(m_1\,\xi+m_2\,\varphi\)}{\cosh\frac{w}{2}},
-\varepsilon\,\tanh\frac{w}{2}\)\qquad\qquad \beta\,e^2=1
\lab{bogogensolnstatic}
\ee

\section{The Hopf topological charge}
\label{sec:hopf}
\setcounter{equation}{0}

We have calculated time dependent and static solutions for the theory
\rf{action}. At each instant of time the solutions define a mapping
from the spatial $\IR^3$ to the target space $S^2$. In fact, all the
solutions we have calculated have finite energy and so the fields go
to a constant at spatial infinity. Therefore, for topological
considerations we can identify the points at infinity and consider the
maps $S^3\rightarrow S^2$, which are classified into homotopy classes
labeled by the integer Hopf index $Q_H$ \cite{bott,afz}. In order to
calculate such index we first consider the mapping of the spatial
$\IR^3$ into a three dimensional sphere $S^3_Z$ parametrized by two
complex functions $Z_{\alpha}$, $\alpha =1,2$, such that $\mid
Z_1\mid^2+\mid Z_2\mid^2=1$. We choose those complex functions, is
terms of the ansatz functions given in \rf{gthetadef} and
\rf{mink-ansatz}, as 
\be
Z_1= \sqrt{1-g}\, e^{i\,m_1\,\xi}\qquad\qquad\qquad 
Z_2= \sqrt{g}\, e^{-i\,m_2\,\vp-i\,m_2\,\zeta}
\ee
which defines the map $\IR^3\rightarrow S^3_Z$. The map
$S^3_Z\rightarrow S^2$ is then given by $u=Z_1/Z_2$, where $u$
parametrizes the plane where the stereographic projection of $S^2$ has
been performed according to \rf{ndef}. The Hopf index is then given by
\cite{bott} 
\be 
Q_H=\frac{1}{4\,\pi^2}\,\int d^3x\,\sum_{i,j,k=1}^3
 \varepsilon_{ijk}\, A_i\,\partial_j\,A_k
\lab{hopfindex}
\ee
where
\be
A_j=
\frac{i}{2}\,\sum_{\alpha=1}^2\,\left[Z_{\alpha}^*\,\partial_j\,Z_{\alpha}-   
Z_{\alpha}\,\partial_j\,Z_{\alpha}^*\right]
\ee
The integral \rf{hopfindex} is performed for a fixed value of time
(i.e. $x^0$), and $Z_{\alpha}$ are functions of $\(\zeta, z, \xi,
\vp\)$. Therefore, in order to perform that integral we follow the
procedure of the appendix \ref{sec:slices} and  eliminate the
coordinate $\zeta$ in favor of the 
time. Then, $\zeta$ becomes a function of $\(x^0,z,\xi\)$ given
by \rf{coszetadef}. Notice that the product of the integrand by the volume
element in \rf{hopfindex} is invariant under change of coordinates. We
then choose the coordinates $\(z, \xi, \vp\)$ to perform the
calculation. The result is 
\be
d^3x\,\sum_{i,j,k=1}^3
 \varepsilon_{ijk}\, A_i\,\partial_j\,A_k = dz\, d\xi\, d\vp\, \(-
 m_1\, m_2\)\, \partial_z g
\ee
and consequently
\be
Q_H = m_1\, m_2\, \left[ g\(0\)-g\(1\)\right]
\lab{hopfindexansatz}
\ee

\section{The energy}
\label{sec:energy}
\setcounter{equation}{0}

According to the discussion in \rf{3ways}-\rf{fphidef} the Hamiltonian
density of the theory \rf{action} is given by
\be
{\cal H}= \frac{\(\beta\, e^2-1\)}{e^2}\, {\cal H}_1 + \frac{1}{e^2}\,
{\cal H}_2
\lab{hamiltoniandensity}
\ee
with ${\cal H}_1$ and ${\cal H}_2$ being given in \rf{h1} and \rf{h2}
respectively. 

The time dependent solutions \rf{bogogensoln} and the static solutions
\rf{bogogensolnstatic} have 
vanishing energy since they both satisfy the constraint \rf{bogomolny}, which
means that ${\cal H}_2$ vanishes. In addition they are solutions only at the
critical point $\beta \, e^2=1$, which means that the first term in
\rf{hamiltoniandensity} vanishes.  

The time dependent solutions \rf{bogosoln-1} have ill defined
energy. The reason 
is that they satisfy the constraint \rf{bogomolnycomplex} and so
${\cal H}_1$ vanishes. However, they are solutions only in the limit
$\beta\, e^2\rightarrow \infty$. So, no matter what ${\cal H}_2$ is
when evaluated on that solution, the first term in
\rf{hamiltoniandensity} is ill defined.  In addition, such solution
have vanishing Hopf index according to \rf{hopfindexansatz}.

There remains to analyze the energy for the time dependent solutions
\rf{nicesolforn}. Since they satisfy \rf{bogomolny} then ${\cal H}_2$
vanishes when evaluated on them. We have now to integrate the first
term in \rf{hamiltoniandensity} on the three dimensional space and
show that the energies of such solutions are indeed time
independent. We use the procedure of appendix \ref{sec:slices} to
eliminate the coordinate $\zeta$ in favor of the time $x^0$, and then
pick up the surfaces of constant time. 

The quantity ${\cal H}_1$ given in \rf{h1} can be written as
\be
{\cal H}_1=32\, \frac{\mid {\dot u}\mid^2}{\(1+\mid u\mid^2\)^2}\, 
\frac{\partial_{\mu}u\,\partial^{\mu}u^*}{\(1+\mid u\mid^2\)^2}-
8\,\frac{\(\partial_{\mu}u\,\partial^{\mu}u^*\)^2}{\(1+\mid u\mid^2\)^4}
\ee
Using the ansatz defined in \rf{gthetadef} and \rf{mink-ansatz}, and
the coordinates introduced in \rf{minkcoord}, one can check that for
the solutions \rf{nicesolforn} one has 
\be
\frac{\partial_{\mu}u\,\partial^{\mu}u^*}{\(1+\mid u\mid^2\)^2} =
-2\,\(\frac{p}{r_0}\)^2 
\ee
with $p$ and $r_0$ defined in \rf{minkcoord}. Using the derivatives
w.r.t. time \rf{invdertau} of
the coordinates introduced in \rf{minkcoord}, one finds that for the
solutions  \rf{nicesolforn} one has 
\be
\frac{\mid {\dot u}\mid^2}{\(1+\mid u\mid^2\)^2}=\(\frac{p}{r_0}\)^2\,
z\, \tau^2
\ee
where $\tau$ is a dimensionless time introduced in
\rf{dimensionlesstime}. Therefore, the quantity ${\cal H}_1$ evaluated on the
solutions  \rf{nicesolforn} is given by
\be
{\cal H}_1= -32\,\(\frac{p}{r_0}\)^4\,\( 1+2\,z\,\tau^2\)
\ee
Using the expression for three dimensional volume element given in
\rf{volumetfixed} one gets that the energies for the solutions
\rf{nicesolforn} are
\br
E&=& \frac{\(\beta\, e^2-1\)}{e^2}\, \int d^3 x\, {\cal H}_1
\nonumber\\
&=& -\frac{16}{r_0\(1+\tau^2\)}\,\frac{\(\beta\, e^2-1\)}{e^2}\,\int
dz\, d\xi\, d\vp\, \( 1+2\,z\,\tau^2\)\, \left[
1- \frac{\(\pm\,\sigma\)}{\sqrt{1+ \tau^2\(1-\sigma^2\)}}\right]
\nonumber
\er
with $\sigma$ as in \rf{coszetadef}, i.e. $\sigma
=\sqrt{1-z}\;\cos\xi$. Notice that the $\sigma$-dependent term in the
integral is odd under the transformation $\xi \rightarrow
\xi+\pi$. Therefore it vanishes when integrated in $\xi$ from $0$ to
$2\,\pi$, and we then obtain that the energy for the solutions
\rf{nicesolforn} are 
\be
E = -\frac{64\,\pi^2}{r_0}\, \frac{\(\beta\, e^2-1\)}{e^2}
\lab{niceenergy}
\ee
and so it is positive for $\frac{\(\beta\, e^2-1\)}{e^2}<0$, in
agreement with Theorem 1 at the end of section \ref{sec:intro}. Notice 
that, similarly to solutions \rf{bogogensoln} and
\rf{bogogensolnstatic}, the solutions \rf{nicesolforn} have vanishing
energy at the critical point $\beta\, e^2=1$.

\section{Other cases}
\label{sec:other}
\setcounter{equation}{0}

The theory \rf{action} on a four dimensional Euclidean
space is  invariant under the conformal group $SO(5,1)$ which
also possesses, like $SO(4,2)$ in Minkowski space, three commuting
$U(1)$ subgroups. Therefore, one can implement a similar ansatz
invariant under the diagonal subgroups of the tensor product of those
$U(1)$ subgroups with the internal $U(1)$ group of phase
transformations $u\rightarrow e^{i\alpha}\, u$. The relevant
coordinates in that case are 
\br
x^0&=& \frac{r_0}{q}\, \sinh y \qquad \qquad \qquad 
x^1= \frac{r_0}{q}\, \sqrt{z}\; \cos\varphi \nonumber\\
x^3&=& \frac{r_0}{q}\, \sqrt{1-z}\; \sin\xi \qquad \quad\;
x^2= \frac{r_0}{q}\, \sqrt{z}\; \sin\varphi 
\lab{euclideancoord}
\er
where
\be
q=\cosh y-\sqrt{1-z}\;\cos\xi 
\ee
The coordinates $z$, $\xi$ and $\varphi$ are the same as in
\rf{minkcoord}, but 
$\zeta$ is replaced by $y$ which can take any real value, $-\infty
\leq y \leq \infty$.
The Euclidean metric in such coordinates becomes
\br
ds^2 = \(\frac{r_0}{q}\)^2\;\left[ dy^2 +
  \frac{dz^2}{4\,z\,\(1-z\)}+\(1-z\)\,d\xi^2 + z\, d\varphi^2\right]
\nonumber
\er 
The ansatz in the Euclidean case is very similar to the Minkowski one,
and it is given by \rf{gthetadef} with the functions given by
\be
g\equiv g\(z\) \qquad \qquad \theta\equiv m_1\, \xi +m_2\, \varphi+\omega\,
y
\lab{euclid-ansatz}
\ee
with $\omega$ being a real parameter (frequency). 

If one replaces the ansatz \rf{gthetadef} and \rf{euclid-ansatz} into the
equations of motion \rf{eom}  associated to \rf{action}, one gets 
that the equation for $\theta$ is automatically satisfied and 
that the profile function $g\(z\)$ has to satisfy \rf{geq} with
$\gamma$ being the same as in \rf{gammadef} and with $\Lambda_M$ replaced by
\be
\Lambda_E \equiv z\,m_1^2+\(1-z\)\,m_2^2+z\(1-z\)\,\omega^2
\lab{lambdaedef}
\ee
In addition if one evaluates the constraints \rf{bogomolny} and
\rf{bogomolnycomplex} one obtains the same relations as in \rf{bogomolny2} and
\rf{bogomolnycomplex2} with $\Lambda_M$ replaced by $\Lambda_E$. Since
$\Lambda_E$ is non-negative on the interval $0\leq z\leq 1$ it follows
that there no non-trivial solutions for the constraint \rf{bogomolnycomplex}
within the ansatz   \rf{gthetadef} and \rf{euclid-ansatz}. 

If $g$ satisfies the constraint \rf{bogomolny}, i.e. \rf{bogomolny2}
with $\Lambda_M$ replaced by $\Lambda_E$, then \rf{geq} in this case
reduces to 
\be
\(\gamma+1\)\partial_z\left[\Lambda_E\,\partial_z g\right]=0
\ee
Performing an analysis similar to that below \rf{geqreduced} one
concludes that \rf{nicesolforn} is a solution in the Euclidean case
with $\varepsilon_i=\pm 1$, $i=1,2,3$, $m_1=\varepsilon_1$,
$m_2=\varepsilon_2$, $\omega =0$, and $\varepsilon_3=1$ for $g=z$ and
$\varepsilon_3=-1$ for $g=1-z$. 

If $\beta\, e^2=1$, i.e. $\gamma =-1$, then $g$ has to satisfy \rf{bogomolny2}
with $\Lambda_M$ replaced by $\Lambda_E$. Performing and analysis
similar to that leading to \rf{bogogensoln} one concludes that
\rf{bogogensoln} is a solution in the Euclidean case with $m_3$
replaced by $\omega$, the coordinate $\zeta$ replaced by $y$, and $w$
being given by \rf{wdef} with $\Lambda_M$ replaced by $\Lambda_E$. In
addition, since $\Lambda_E$ is non-negative on the interval $0\leq
z\leq 1$, for any values of $m_1$, $m_2$ and $\omega$, one does not
have restrictions similar to those  in  \rf{restrictions} involving
$m_i$, $i=1,2,3$.

One can also consider, in Minkowski space-time, a modification of the
theory \rf{action} given by 
\be
{\tilde {\cal L}} = -\frac{1}{e^2} \, H_{\mu\nu}^2 + \frac{\beta}{2}\,
\mid \left(\partial_{\mu} {\vec n}\right)^2\mid \, 
\left(\partial_{\nu} {\vec n}\right)^2
\lab{actionposdef}
\ee
It has the same target space $SO(3)$ and conformal $SO(4,2)$ symmetries as 
\rf{action}, and so the ansatz  \rf{gthetadef} and \rf{mink-ansatz} 
also works here. It has the advantage  of having a  positive
definite Hamiltonian for $\beta >0$ and $e^2>0$ \cite{adampos}.  

If one replaces the ansatz  \rf{gthetadef} and \rf{mink-ansatz} into
the Euler-Lagrange equations associated to \rf{actionposdef}, one
obtains that the equation for the field $\theta$ is automatically
satisfied and the equation for $g$ leads to \rf{geq} with $\Lambda_M$
being given by \rf{lambdadef}, but with the parameter $\gamma$ being
now given by 
\be
\gamma =\(1+\frac{2}{\beta\,e^2}\)
\ee
Therefore all the solutions we have constructed for \rf{action} are
also solutions of \rf{actionposdef}. In fact, the only point we have
to pay attention is that the critical point now corresponds to
$\beta\,e^2=-1$, instead of $\beta\,e^2=1$. Therefore, 
\rf{nicesolforn} are solutions of \rf{actionposdef} for any value of
$\beta$ and $e^2$. In addition \rf{bogogensoln} as well as the static
configurations \rf{bogogensolnstatic} are solutions of
\rf{actionposdef} for $\beta\,e^2=-1$. The configurations
\rf{bogosoln-1} are also solutions of \rf{actionposdef} for
$\beta\,e^2\rightarrow \infty$.

\section{Summary and conclusions}
\label{sec:conclusions}
\setcounter{equation}{0}

We have constructed exact solutions for the theory \rf{action}
exploring its conformal symmetry $SO(4,2)$ and the target space
symmetry $u\rightarrow e^{i\,\alpha}\, u$. Using those symmetries we
built an ansatz invariant under the diagonal subgroup of the direct
product of the $U(1)$ target space group of phase transformations with
three commuting $U(1)$ subgroups of $SO(4,2)$. The ansatz is given by
\rf{gthetadef} and \rf{mink-ansatz} and is based on the toroidal like
coordinates \rf{minkcoord}. The crucial point of our construction is
that our solutions are in fact solutions of Bogomolny type equations
given by the \rf{bogomolny} and \rf{bogomolnycomplex}. Those two
equations define submodels of the theory \rf{action} possessing an
infinite number of local conserved currents given by \rf{conscurr} and
\rf{conscurr2}. In addition there exist critical points, defined by
special values of the coupling constants, where the theory \rf{action}
becomes even more integrable. The first order differential equation
\rf{bogomolny} together with the condition $\beta\,e^2=1$, imply the
second order equations of motion of the theory \rf{action}. The same
is true for the equation \rf{bogomolnycomplex} and the condition
$\beta\,e^2\rightarrow \infty$. The exact solutions we have
constructed are the following:
\begin{enumerate}
\item The configurations \rf{nicesolforn} are solutions of
  \rf{bogomolny} as well as of the equations of motion of \rf{action},
  for any value of the coupling constants $\beta$ and $e^2$. Their Hopf
  topological charge, according to \rf{hopfindexansatz}, are given by
  $Q_H= -\varepsilon_1\,\varepsilon_2\,\varepsilon_3$, with
  $\varepsilon_i=\pm 1$, $i=1,2,3$. All those eight solutions have the
  same energy and given by \rf{niceenergy}.
\item The configurations \rf{bogogensoln} are solutions of
  \rf{bogomolny}, and for $\beta\,e^2=1$ are also solutions of  the
  equations of motion of \rf{action}. They all have vanishing energies,
  and their Hopf charges are $Q_H= -\varepsilon\,m_1\, m_2$, with
  $\varepsilon=\pm 1$, and $m_i$, $i=1,2$, any integers (see
  \rf{hopfindexansatz}). Notice that such solutions have a free
  parameter $\kappa$ inside the definition of $w$ given in \rf{wdef},
  that allows the deformation of the surfaces of constant $n_3$,
  i.e. the third component of the triplet ${\vec n}$.  
\item The static configurations \rf{bogogensolnstatic} are solutions
  of \rf{bogomolny}, and for $\beta\,e^2=1$ are also solutions of  the
  equations of motion of \rf{action}. They all have vanishing energies,
  and their Hopf charges are $Q_H= -\varepsilon\,m_1\, m_2$, with
  $\varepsilon=\pm 1$, and $m_i$, $i=1,2$, any integers. They also
  carry the same free parameter $\kappa$ and are the static version of
  the solutions \rf{bogogensoln}. 
\item The configurations \rf{bogosoln-1} are solutions of
  \rf{bogomolnycomplex}, and for $\beta\,e^2\rightarrow \infty$ are also
  solutions of  the equations of motion of \rf{action}. Such solutions
  are not very interesting because they have vanishing Hopf
  topological charge, and their energies are ill defined. In this
  sense, the condition \rf{bogomolnycomplex} does not lead to relevant
  solutions within the ansatz  \rf{gthetadef} and \rf{mink-ansatz}. 
\end{enumerate}

If one wants to find solutions of the theory \rf{action} outside the
integrable sub-sector defined by the condition \rf{bogomolny} (or
\rf{bogomolnycomplex}), one has to solve the ordinary differential
equation \rf{geq}, for the profile function $g\(z\)$, using numerical
methods. Even though that is quite feasible we do not pursue it in
this paper. Perhaps the main difficulties facing the numerical
integration of \rf{geq} are related to its critical points at $z=0$
and $z=1$. The best strategy would perhaps be to use the shooting
method starting from both ends at $z=0$ and $z=1$ and try to adjust the
initial conditions  to match the two branches around $z=1/2$.   

The theory \rf{action} is closed related to the so-called extended
Skyrme-Faddeev model defined by the Lagrangian 
\be
{\cal L}_{\rm ESF} = M^2\, \partial_{\mu} {\vec n}\cdot\partial^{\mu} {\vec n}
 -\frac{1}{e^2} \, \(\partial_{\mu}{\vec n} \wedge 
\partial_{\nu}{\vec n}\)^2 + \frac{\beta}{2}\,
\left(\partial_{\mu} {\vec n}\cdot\partial^{\mu} {\vec n}\right)^2
\lab{lagrangian}
\ee
In \cite{vortexlaf} it was constructed exact vortex solutions for
\rf{lagrangian}. Those vortices (static and time-dependent) are
solutions of the constraint \rf{bogomolny} and solve the equations of
motion only on the critical point $\beta\, e^2=1$. In that sense they
are self-dual vortices. In \cite{fst} numerical
solutions for hopfions were constructed for the theory
\rf{lagrangian}. Those solutions have some interesting
properties. They are close to satisfy the constraint \rf{bogomolny},
and they become solutions of it only in the limit $\beta\, e^2\rightarrow
1$ (the point $\beta\, e^2=1$ was not accessible
numerically). In addition, as $\beta\,e^2$ approaches unity, the quantity
$a^2 = -e^2\,r_0^2\,M^2$ approaches zero.  That can be interpreted by
the fact that the size of the solution (determined by $r_0$)
approaches zero and so the solution collapses. However, one can
connect that fact to the solutions constructed in this paper. As $a^2$
approaches zero one can think of $M^2$ going to zero and the theory
\rf{lagrangian} reducing to \rf{action}. Therefore, the numerical
solutions of \cite{fst} should reduce to the self-dual hopfions
constructed here, in the limit $\beta\, e^2\rightarrow 1$.

\newpage

\appendix 

\section{Analysis of $\Lambda_M$}
\label{sec:appendixlambdam}
\setcounter{equation}{0}

Notice that for $m_3=0$, we have that $\Lambda_M$ given in
\rf{lambdadef} is non-negative on the interval $0\leq z\leq 1$. In
addition, for $m_1=m_2=0$, we have that $\Lambda_M$ is non-positive on
the same interval. We also have  that
\be
\Lambda_M\mid_{z=0}=m_2^2\geq 0\; , \qquad\qquad 
\Lambda_M\mid_{z=1}=m_1^2\geq 0
\ee
For
$m_3\neq 0$ we write
\be
\Lambda_M  = m_3^2\(z-z_{+}\)\(z-z_{-}\)\; , \qquad \qquad 
z_{\pm}=\frac{1}{2}\(-b\pm \sqrt{\Delta}\)
\ee
with
\be
b= r_{+}\,r_{-}-1\; ,\qquad \quad\Delta=\(r_{+}^2-1\)\(r_{-}^2-1\)\; ,
\qquad \quad r_{\pm}=\frac{m_1\pm m_2}{m_3}
\ee
For $r_{+}^2>1$ and $r_{-}^2<1$ or $r_{+}^2<1$ and $r_{-}^2>1$, we
have $\Delta <0$ and so $\Lambda_M$ does not have real zeros, and
so  $\Lambda_M\geq 0$ in that case. 

Notice we can write $b$ as
\be
b= \frac{1}{2}\left[ \(r_{+}-1\)\(r_{-}+1\)+\(r_{+}+1\)\(r_{-}-1\)\right]
\ee
Therefore, for $r_{\pm}>1$ or $r_{\pm}<-1$ we can write
\be 
z_{\pm}= -\frac{1}{4}\left[\sqrt{\(r_{+}-1\)\(r_{-}+1\)}\mp 
\sqrt{\(r_{+}+1\)\(r_{-}-1\)}\right]^2
\ee
Consequently, $z_{-}< 0$, and $z_{+}\leq 0$ (with $z_{+}=0$ for
$r_{+}=r_{-}$, or $m_2=0$), and 
$\Lambda_M$ has no zeros on the interval $0< z\leq 1$, and so
$\Lambda_M\geq 0$ on that interval. 
 
For $r_{+}>1$ and $r_{-}<-1$,  we write
\br
z_{\pm}&=& \frac{1}{4}\left[\sqrt{\(r_{+}+1\)\(1-r_{-}\)}\pm
\sqrt{\(1-r_{+}\)\(r_{-}+1\)}\right]^2\nonumber\\
&=&\frac{1}{4}\left[\sqrt{4+2\,s_{+}+2\, s_{-}+s_{+}\,s_{-}}\pm
\sqrt{s_{+}\,s_{-}}\right]^2
\lab{nicecase}
\er 
with $s_{+}=r_{+}-1>0$, and $s_{-}=-\(1+r_{-}\)>0$. Consequently,
$z_{+}> 1$, and $z_{-}\geq 1$ (with $z_{-}=1$ for $r_{+}=-r_{-}$, or
$m_1=0$), and $\Lambda_M$ 
has no zeros on the interval $0\leq z< 1$, and so  $\Lambda_M\geq 0$
on that interval. 
For $r_{+}<-1$ and $r_{-}>1$, we use the same reasoning as in
 \rf{nicecase}, with the replacement $r_{+}\leftrightarrow r_{-}$, and
 the fact the $b$ and $\Delta$ are invariant under such interchange. 

For $-1\leq r_{\pm}\leq 1$ we write
$$
z_{\pm}= \frac{1}{4}\left[\sqrt{\(r_{+}+1\)\(1-r_{-}\)}\pm 
\sqrt{\(1-r_{+}\)\(r_{-}+1\)}\right]^2=\left[\sqrt{\(1-t_{+}\)\(1-t_{-}\)}\pm
  \sqrt{t_{+}\,t_{-}}\right]^2 
$$
with $t_{+}=\(1-r_{+}\)/2$, and $t_{-}=\(r_{-}+1\)/2$, and so $0\leq
t_{\pm}\leq 1$. Consequently we have  $0\leq z_{\pm}\leq 1$, and so in that
case, $\Lambda_M$ does change sign on the interval $0\leq z\leq 1$. 
 
Summarizing we have that, on the interval $0\leq z\leq 1$, 
\br
\Lambda_M = \left\{\begin{array}{l}
\mbox{\rm has no definite sign for $\(\mid m_1\mid + \mid
  m_2\mid\)^2\leq m_3^2$}\\
\mbox{\rm is non-positive for $m_1=m_2=0$} \\
\mbox{\rm is non-negative otherwise}\\
\end{array}\right.
\er
 
Notice that $\Lambda_E$ introduced in \rf{lambdaedef} is non-negative
on the interval  $0\leq z\leq 1$, as long as $\omega$ is taken to be 
real. However, if one allows it to be pure imaginary,
i.e. $\omega=i\,{\tilde \omega}$, then the analysis above applies
equally well for $\Lambda_E$ with the replacement $m_3\leftrightarrow {\tilde
  \omega}$.

\section{The surfaces of constant time}
\label{sec:slices}
\setcounter{equation}{0}

In order to evaluate the Hopf charges and the energies of the solutions
we need to evaluate integrals over the three dimensional space at a
given fixed value of time. The coordinates introduced in
\rf{minkcoord} are such that none of them are parallel to
time. Therefore we shall choose to eliminate the coordinate $\zeta$ in
favor of time and go from $\(\zeta, z, \xi, \vp\)$ to $\(x^0, z, \xi,
\vp\)$, which is then a non-orthogonal system of coordinates. In order
to express $\zeta$ in terms of the new coordinates we use the first
relation in \rf{minkcoord}, which leads to  
\be
\tau^2\, p^2 = 1- \cos^2 \zeta
\ee
where we have introduced the dimensionless time
\be
\tau \equiv \frac{x^0}{a}=\frac{c\, t}{a}
\lab{dimensionlesstime}
\ee
That is a quadractic relation for $\cos\zeta$. Solving it one has
\be
\cos \zeta_{\pm} = \frac{\tau^2\, \sigma \pm \sqrt{1+ 
    \tau^2\left(1 -\sigma^2\right)}}{1+\tau^2} \qquad\qquad {\rm with}
\qquad\qquad 
\sigma\equiv \sqrt{1-z}\, \cos\xi
\lab{coszetadef}
\ee 
with both signs giving equivalent relations among $\zeta$, $\tau$, $z$
and $\xi$. In fact, $\cos \zeta_{+}\(\sigma\)=-\cos\zeta_{-}\(-\sigma\)$. \\
The three dimensional spatial volume element is then given by
\be
dx^1\,dx^2\,dx^3=
\(\frac{r_0}{p}\)^4 \,\frac{1}{2\,r_0}\;\frac{1}{1+\tau^2}\;\left[
1- \frac{\(\pm\,\sigma\)}{\sqrt{1+ \tau^2\(1-\sigma^2\)}}\right] \;dz\, d\xi\,
d\varphi
\lab{volumetfixed}
\ee
with the signs in front of $\sigma$ being the same as those in
\rf{coszetadef}. 
In addition, the derivatives of the coordinates with respect to
time, useful in the calculation of the energy, are given 
by 
\br
\frac{\partial \zeta}{\partial x^0}&=& 
\pm\,\(\frac{p}{r_0}\)\, \sqrt{1+ \tau^2\(1-\sigma^2\)}
\nonumber\\
\frac{\partial z}{\partial x^0}&=& 
\(\frac{p}{r_0}\)\, 2\, z\, \sqrt{1-z}\, \cos\xi \;\;\tau 
\nonumber\\
\frac{\partial \xi}{\partial x^0}&=& 
\(\frac{p}{r_0}\)\,\frac{\sin\xi}{\sqrt{1-z}}\;\;\tau
\nonumber\\
\frac{\partial \varphi}{\partial x^0}&=&0
\lab{invdertau}
\er

\vspace{3cm}

\noindent {\bf Acknowledgments:} LAF is grateful to N. Sawado and
K. Toda for helpful discussions. LAF is partially supported by CNPq.


\newpage

\begin{thebibliography}{99}
\bibitem{coleman}
S.~R.~Coleman,
  ``Quantum sine-Gordon equation as the massive Thirring model,''
  Phys.\ Rev.\  D {\bf 11}, 2088 (1975).\\
S.~Mandelstam,
  ``Soliton operators for the quantized sine-Gordon equation,''
  Phys.\ Rev.\  D {\bf 11}, 3026 (1975).

\bibitem{duality}
C.~Montonen and D.~I.~Olive,
  ``Magnetic Monopoles As Gauge Particles?,''
  Phys.\ Lett.\  B {\bf 72}, 117 (1977).\\
C.~Vafa and E.~Witten,
  ``A Strong coupling test of S duality,''
  Nucl.\ Phys.\  B {\bf 431}, 3 (1994)
  [arXiv:hep-th/9408074].\\
N.~Seiberg and E.~Witten,
  ``Electric - magnetic duality, monopole condensation, and confinement in N=2
  supersymmetric Yang-Mills theory,''
  Nucl.\ Phys.\  B {\bf 426}, 19 (1994)
  [Erratum-ibid.\  B {\bf 430}, 485 (1994)]
  [arXiv:hep-th/9407087].

\bibitem{lax}
 P.~D.~Lax,
  ``Integrals Of Nonlinear Equations Of Evolution And Solitary Waves,''
  Commun.\ Pure Appl.\ Math.\  {\bf 21}, 467 (1968).
V.E. Zakharov and A.B. Shabat, {\it Zh. Exp. Teor. Fiz. } {\bf 61} 
(1971) 118-134;  english transl. {\it Soviet Phys. JETP} {\bf 34} (1972) 62-69.

\bibitem{afs}
 O. Alvarez, L.A. Ferreira, J. Sanchez Guillen, ``A
  new approach to 
integrable theories in any dimension'', 
{\em Nucl. Phys.} {\bf B529} (1998) 689-736, [arXiv:hep-th/9710147]

\bibitem{afs-review}
  O.~Alvarez, L.~A.~Ferreira and J.~Sanchez-Guillen,
  ``Integrable theories and loop spaces: fundamentals, applications and new
  developments,''
  Int.\ J.\ Mod.\ Phys.\  A {\bf 24}, 1825 (2009)
  [arXiv:0901.1654 [hep-th]].

\bibitem{skyrme}
T.~H.~R.~Skyrme,
  ``A Nonlinear field theory,''
  Proc.\ Roy.\ Soc.\ Lond.\  A {\bf 260}, 127 (1961).

\bibitem{faddeev-niemi}
L.~D.~Faddeev and A.~J.~Niemi,
  ``Knots and particles,''
  Nature {\bf 387}, 58 (1997)
  [arXiv:hep-th/9610193]; 

\bibitem{mantonbook}
N.~S.~Manton and P.~Sutcliffe,
  ``Topological solitons,''
{\it  Cambridge, UK: Univ. Pr. (2004) 493 p}

\bibitem{babelon}
O.~Babelon and L.~A.~Ferreira,
  ``Integrability and conformal symmetry in higher dimensions: A model with
  exact Hopfion solutions,''
  JHEP {\bf 0211}, 020 (2002)
  [arXiv:hep-th/0210154]; 

\bibitem{afz}
 H. Aratyn, L.A. Ferreira, A.H. Zimerman,
``Exact  static soliton solutions  
of $3+1$ dimensional integrable theory with nonzero Hopf numbers,''
{\em Phys. Rev. Lett.} {\bf 83} (1999) 1723-1726,  [arXiv:hep-th/9905079]

\bibitem{laf1}
L.~A.~Ferreira,
  ``Exact time dependent Hopf solitons in 3+1 dimensions,''
  JHEP {\bf 0603}, 075 (2006)
  [arXiv:hep-th/0601235].

\bibitem{laf2}
L.~A.~Ferreira,
  ``Euclidean 4d exact solitons in a Skyrme type model,''
  Phys.\ Lett.\  B {\bf 606}, 417 (2005)
  [arXiv:hep-th/0406227].

\bibitem{gies}
H.~Gies, 
``Wilsonian effective action for SU(2) Yang-Mills theory with
Cho-Faddeev-Niemi-Shabanov decomposition,''
Phys.\ Rev.\ D {\bf 63}, 125023 (2001), hep-th/0102026

\bibitem{chofn}
L.~D.~Faddeev and A.~J.~Niemi,
  ``Partially dual variables in SU(2) Yang-Mills theory,''
  Phys.\ Rev.\ Lett.\  {\bf 82}, 1624 (1999)
  [arXiv:hep-th/9807069]; 
Y.~M.~Cho,
``A Restricted Gauge Theory,''
Phys.\ Rev.\ D {\bf 21}, 1080 (1980); and 
``Extended Gauge Theory And Its Mass Spectrum,''
Phys.\ Rev.\ D {\bf 23}, 2415 (1981).

\bibitem{vortexlaf} 
  L.~A.~Ferreira,
  ``Exact vortex solutions in an extended Skyrme-Faddeev model,''
{\em Journal of High Energy Physics} {\bf JHEP05(2009)001}, 
  arXiv:0809.4303 [hep-th].

\bibitem{fst}
L.~A.~Ferreira, N.~Sawado and K.~Toda,
  ``Static Hopfions in the extended Skyrme-Faddeev model,''
  {\em Journal of High Energy Physics} JHEP {\bf 0911}, 124 (2009)
  [arXiv:0908.3672 [hep-th]].

\bibitem{sutcliffe}
 R.~A.~Battye and P.~M.~Sutcliffe,
  ``Knots as stable soliton solutions in a three-dimensional classical  field
  theory,''
  Phys.\ Rev.\ Lett.\  {\bf 81}, 4798 (1998)
  [arXiv:hep-th/9808129]; 

\bibitem{hietarinta}
J.~Hietarinta, P.~Salo,
  ``Ground state in the Faddeev-Skyrme model,''
  Phys.\ Rev.\ D {\bf 62}, 081701 (2000).

\bibitem{derrick}
  G.~H.~Derrick,
  ``Comments on nonlinear wave equations as models for elementary particles,''
  J.\ Math.\ Phys.\  {\bf 5}, 1252 (1964).

\bibitem{bott} R. Bott and L.W. Tu, 
{\em Differential Forms in Algebraic Topology} (Graduate Texts in
Mathematics: 82), Springer 1982. 

\bibitem{adampos}
C.~Adam, J.~Sanchez-Guillen and A.~Wereszczynski,
  ``k-defects as compactons,''
  J.\ Phys.\ A  {\bf 40}, 13625 (2007)
  [arXiv:0705.3554 [hep-th]].\\
C.~Adam, N.~Grandi, J.~Sanchez-Guillen and A.~Wereszczynski,
  ``K fields, compactons, and thick branes,''
  J.\ Phys.\ A  {\bf 41}, 212004 (2008)
  [arXiv:0711.3550 [hep-th]].

\end{thebibliography}
\end{document}